\documentclass{appolb}
\usepackage{graphics}
\begin{document}
\pagestyle{empty}
% \eqsec  % uncomment this line to get equations numbered by (sec.num)
\title{SPIN-ORBIT PENDULUM -- RELATIVISTIC EXTENSION\thanks
{Presented at XXVI Mazurian Lakes School of Physics on 
 Nuclear Physics at the Turn of the Century, Krzyze, Poland,
 September 1-11, 1999. }%
% you can use '\\' to break lines
}
\author{\underline{M.\ Turek},  P.\ Rozmej
\address{Instytut Fizyki, Uniwersytet Marii Curie-Sk\l odowskiej, \\ 
Pl.\ M. Curie-Sk\l odowskiej~1, 20-031 Lublin, Poland}
\and
R.\ Arvieu
\address{Institut des Sciences Nucl\'eaires, F-38026 Grenoble-Cedex, 
France}
}

\maketitle
\begin{abstract}
We discuss an extension of the theory of {\em spin-orbit pendulum}
phenomenon given in \cite{sop}
to relativistic approach. It is done within the so called Dirac
Oscillator. Our first results, focusing on 
circular wave packet motion have been published recently \cite{doc}. 
The scope of this paper is motion of a linear wave packet.
In relativistic approach we found {\em Zitterbewegung} in spin-orbit motion 
(in Dirac representation) due to coupling to
negative energy states. This effect is washed out in the 
Foldy-Wouthuysen representation. Another important change with
respect to non-relativistic case is the loss of periodicity.
The phenomenon reminds the time evolution of population inversion
in Jaynes-Cummings model.        
\end{abstract}
 \PACS{03.65.Sq, 03.65.Ge}  
\section{Introduction} 
Few years ago we investigated motion of a wave packet (WP)
representing a fermion in the harmonic oscillator potential with
spin-orbit coupling. In this model (and non-relativistic approach)
we have found the {\em spin-orbit pendulum} phenomenon 
\cite{sop}. We predicted there that  
spin-orbit forces creates interesting  oscillations of 
expectation values of spin components if particle state is 
prepared as a well localized wave packet. Then expectation values
of the spin and angular momentum components oscillate periodically.
During one period of this time evolution, the spin collapses 
(at the same time the wave function initially pure in both subspaces 
gets maximum entanglement) then (almost) restores in the opposite
direction, collapses again and restores exactly.

The aim of this paper is to extend the model to relativistic WPs,
focusing on cases corresponding to linear classical trajectories
of a  particle (the other interesting case of circular trajectories
is discussed elsewhere \cite{doc}). The relativistic harmonic oscillator
has been introduced many years ago in particle physics  \cite{olddo}
and then refreshed \cite{mosh} under the name {\em Dirac Oscillator} (DO).
More recently the behavior of the WPs in DO was studied in both
Dirac and Foldy-Wouthuysen (F-W) representations in 1+1 dimensions 
\cite{nog}.
\section{Dirac Oscillator} 
DO is described by the time dependent equation:
\begin{equation}\label{DO}
i \hbar { { \partial \Psi} \over { \partial t} } = 
H_{DO} \Psi= c \left[ \, \mbox{\boldmath $ \alpha $} \cdot 
(\, {\mathbf p}- i m \omega 
{\mathbf r} \beta ) + m c \beta \right] \, ,
\end{equation}
where $ \mbox{\boldmath $ \alpha $}, \beta $ are usual Dirac matrices.
One can show that both the large and small component of an DO 
eigenstate are proportional to $|N(ls)j m_j\rangle$ 
-- an eigenstate of the 3d-HO with a spin-orbit coupling.   
The energy spectrum has the form of
\begin{equation}
E_{Nlj}= \pm \, mc^2 \sqrt{1+r\,A_{Nlj}} \hspace{3ex}
A_{Nlj}= \left\{ \matrix{ 2(N-j)+1 & \hbox{ \ for } j = l+{{1} 
\over {2}} \cr
 2(N+j)+3 & \hbox{ \ for } j = l-{{1} \over {2}} \cr } \right.
\end{equation}
The parameter $r=\hbar\omega/m c^2$ enables, 
if it is small enough, a transition to the
non-relativistic limit.
\section{Initial Form of the WP}
We study the evolution of a Gaussian WP which is initially    
centered at $\mathbf r_0 $ and has the average momentum $ \mathbf p_0$ 
(i.e. a 3d-HO coherent state). Moreover, the initial WP is an eigenstate 
of the spin pointed at some arbitrary direction defined by two numbers 
$ \alpha$ and  $ \beta $:
 \begin{equation}
 \label{gauss}
 \Psi({\mathbf r}, t=0)= {{1} \over {(2 \pi)^{ 3 \over 4} \sigma^{3 \over 2} 
 }} \exp{ \left[  {{({{\mathbf r} -{ \mathbf r_0}})^2} 
 \over {2 \sigma^2}} + i 
 {{{{\mathbf p_0} \cdot { \mathbf r}}} \over {\hbar}}
 \right] } \, \pmatrix{ \alpha \cr \beta \cr 0 \cr 0} \, ,
 \end{equation}
 where $ \sigma= \sqrt{\hbar \slash m \omega}$. 
  In the following we present the time evolution of {\em linear} WP
 (\ref{gauss})
 corresponding to the following initial conditions:
${\mathbf r_0} = z_0 \, \hat z$ and 
${\mathbf p_0} = p_0 \, \hat z $. It is possible thanks to decomposition
 \cite{sop} :
\begin{equation} 
\Psi_{\rm lin}(t=0) =\sum_N^{\infty}\sum_{l=0(1)}^{N} \lambda_{Nl} \, 
| N,l,0 \rangle \pmatrix{ \alpha \cr \beta \cr 0 \cr 0} \, .
\end{equation}
\section{Time evolution}
\begin{figure}[t] 
 \resizebox{1\textwidth}{!}
 {\includegraphics{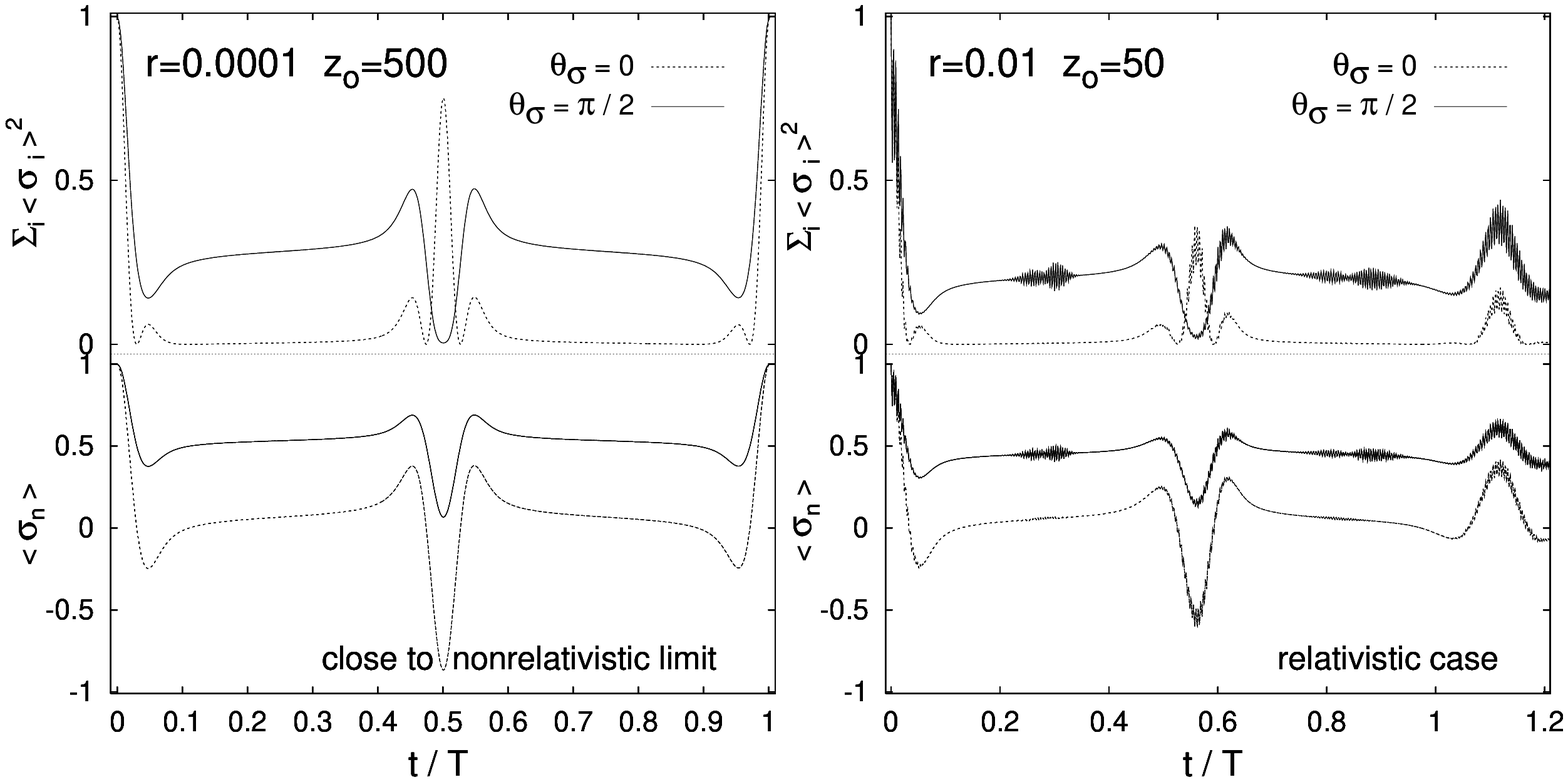} }
\caption{Dependence of  the spin averages time evolution for 
 the linear WP on the initial spin direction. Two cases of $r$ values 
 are presented ($r=0.0001$ and $r=0.01$), $\langle\sigma_n\rangle $ is
 an average value of the spin projection on the initial spin direction.}
\label{f1} %1
\end{figure}

\begin{figure}[t] 
\hspace{10mm} \resizebox{0.7\textwidth}{!}
 {\includegraphics{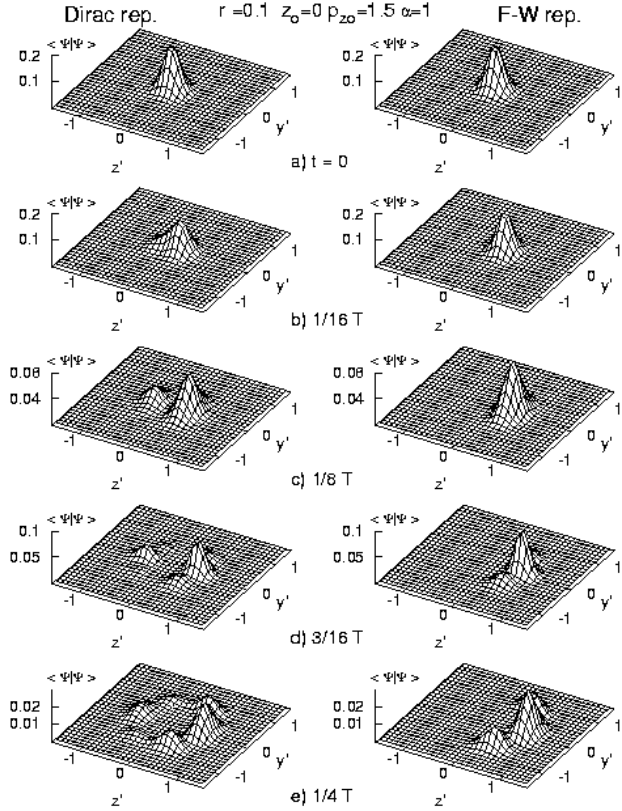} }\vspace{-3mm}
\caption{Comparison of the evolution of the linear WP (r=0.5) in Dirac
 and F-W representations. Initial spin direction is parallel to
 ${\mathbf p_0}$. }
\label{f2} %2
\end{figure}

Fig.\ \ref{f1} shows the behavior of spin averages for the linear WP
in the Dirac representation.
The transition from the pure state of the spin in a well-defined 
direction to the entangled 
state is particularly clear in the case of 
$(\theta_{\sigma}=0)$. The initial conditions are such that WP
is launched along $Oz$ axis from the center with spin parallel to
${\mathbf p_0}$.
Notice the {\em Zitterbewegung} and 
and deviation from periodicity for the relativistic case.
An example of the spatial WP motion is shown in Fig.\ \ref{f2}.
The probability density is shown on $zOy$ plane (there is cylindrical 
symmetry with respect to $Oz$ axis). Sub-packets corresponding to
states of positive and negative energies are well seen in Dirac
representation, whereas only positive energy states are present
in the F-W representation. \\[1mm]

{\sf Acknowledgmnet~} M.T.\ and P.R.\ thank for support of Polish 
Committee for Scientific Research (KBN) under the grant 2 P03B 143 14.
\vspace{-3mm} 
%\newpage

\end{document}